# Secure Federated Learning for Cognitive Radio Sensing

Małgorzata Wasilewska[1,2], Hanna Bogucka[1,2], H. Vincent Poor[3]

[1]Poznan University of Technology, Poland, [2]RIMEDO Labs, Poland; [3]Princeton University, USA

**Abstract**

This paper considers reliable and secure Spectrum Sensing (SS) based on Federated Learning (FL) in the Cognitive Radio (CR) environment. Motivation, architectures, and algorithms of FL in SS are discussed. Security and privacy threats on these algorithms are overviewed, along with possible countermeasures to such attacks. Some illustrative examples are also provided, with design recommendations for FL-based SS in future CRs.

## I. Introduction

The idea of Cognitive Radio (CR) was proposed two decades ago to enhance the operation of radio devices and networks by embedding operational-environment awareness and artificial intelligence in them. It evolved to cognitive use of the spectrum opportunities by dynamic access to temporarily unused frequency bands (so-called *spectrum holes*). In this approach, an unlicensed user (called a *secondary user* or CR user) can detect a spectrum hole and transmit signals in this frequency band provided that interference generated to Primary (licensed) Users' (PUs) communication channels is kept up to the allowable level.

The CR technology's essential element is detecting the spectrum holes, i.e., spectrum resources not occupied by a PU. This may be achieved by getting the relevant information from a dedicated database (if available in a given location), often called a *radio-environment map,* or by real-time measurement of the PU's activity, called Spectrum Sensing (SS). Sensing is a process intended to uncover spectrum occupation and spectrum holes. It allows for taking advantage of spectrum opportunities, dynamic spectrum access, resource management for anticipated traffic, etc. It can be based on various methods, e.g., signal energy or feature detection. Autonomous sensing by each distinct SU has not been considered reliable enough due to the adverse propagation effects of a wireless channel. Cooperative Sensing (CS) in CR networks mitigates these effects. It requires sensors to share their local sensing information with a so-called *fusion center*, which determines the spectrum occupancy in a given area (e.g., based on majority voting). This sharing of information causes CS to be prone to privacy and security attacks [1].

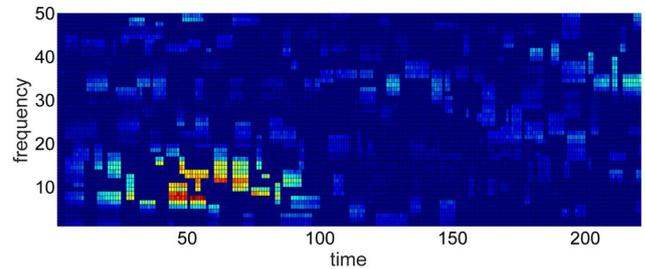

Figure 1. Image representing typical detected signal energy in time and frequency RBs.

Spectrum sensing can be improved by employing Machine Learning (ML). This is because time and frequency Resource Block (RB) allocation for PUs in 4G, 5G, and 6G systems is based on channel qualities, avoids resource fragmentation, and follows time-frequency traffic patterns. Thus, time and frequency fading patterns, as well as RB allocation, expose dependencies that ML can uncover (see Fig. 1). There are several ML algorithms (or structures) that can be used for this purpose, similar to image processing, e.g., neural networks such as Convolutional Neural Networks (CNNs). Based on these uncovered patterns, predictions can also be made regarding future spectrum occupation, which is essential for the efficient operation of CRs and PU transmission protection [2].

ML-based SS using autonomous sensors has limited reliability due to distortions of a wireless channel. However, if frequency-selective fading dependencies can be uncovered (within the channel coherence time), the probability of misdetection can be reduced. Alternatively, a centralized ML approach would require extensive training datasets with high-resolution localization data, which may be impractical to acquire.

To overcome these obstacles, Federated Learning (FL) has been considered for SS ([2],[3]), which enables a group of sensors to execute a common learning task by exchanging their local model parameters (or a distilled representative part of a model) instead of raw data to accomplish aggregate analytics. Thereby, FL is considered to be a privacy-by-design technique while achieving high learning accuracy [4]. Nevertheless, FL applied to wireless systems, including SS, does not guarantee the levels of security required by modern communication systems. FL systems are vulnerable to attacks that target each stage of training and decision-making. Attackers can exploit flaws in FL systems in various ways, e.g., by corrupting the training data or local model updates at CR User Equipment (UE) or intercepting the model updates exchanged with the central server [5]. Thus, although FL enhances privacy, privacy is not guaranteed without further protection [6].

Unlike some recent reviews on FL for CR ([1]-[4],[7]), general-purpose FL algorithms security ([5],[6],[8]-[12]), ML-based resource sharing and wireless security ([13],[14]), or research papers on specific threats and countermeasures (e.g. [15] on FL SS robustness to poisoning-attacks) this article is an overview of reliable and secure FL-based SS. In Section II, we discuss the application of FL for SS. Section III reviews security and privacy attacks on these algorithms. In Section IV, countermeasures to such attacks are considered. We conclude our considerations in Section V.

## II. Federated Learning for Spectrum Sensing and Prediction

As mentioned in the Introduction, the issue with spectrum sensing by individual agents is limited reliability, i.e., the quality of decisions is affected by the agent's specific radio environment, limited computational and memory resources, as well as limited availability of labeled datasets (so there is a need to download them from an external database, which may introduce errors and delay). Federated learning is a concept that can resolve the problem of handling distributed datasets. In this case, all the training data is kept where it is generated (or measured), and only locally trained models are transferred to the central FL server. Such decentralized ML reduces the required radio resources (bandwidth, time, and energy) and the data-processing latency by sending only the model parameters instead of the raw data stream [7].

FL is an iterative procedure employing edge devices (sensors onboard CR UEs, called FL nodes or FL agents) that develop their ML models based on their locally measured data. The local models are then exchanged in a centralized or decentralized way to create one aggregated model, which is shared among the devices. In the case of the centralized approach, locally developed model parameters, either all of them or only the ones that define some part of a model (e.g., distilled models' parameters [3]), are transmitted to an FL server, where the aggregated model is created. FL nodes then adjust the corporate model to their local data, and the process of local training, exchanging models, and aggregating them is repeated (see Fig. 2, where FL nodes are considered CRs). Apart from the model aggregation, the FL server may also be in charge of FL node clustering, reflecting the location-specific availability of spectrum resources.

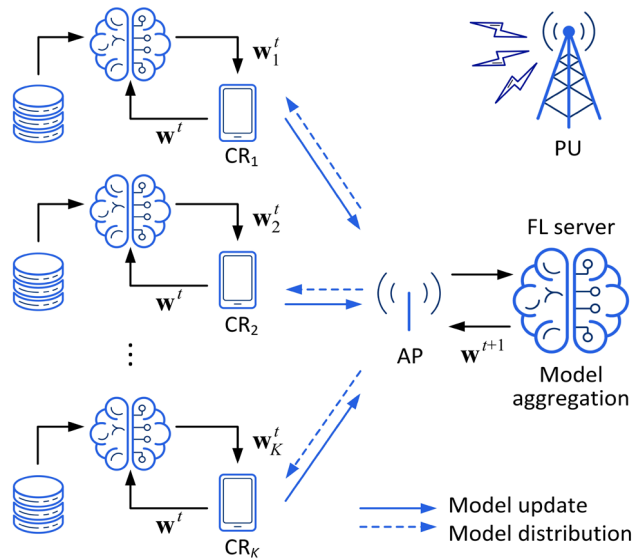

Figure 2. Federated Learning for spectrum sensing.

Apart from the advantages mentioned above, FL-based sensing provides a new incoming CR UE with a spectrum-occupancy-reflecting model suited for its current wireless environment and location without the need to collect data and train a model. It can also adapt to the changing radio environment. Whenever the radio channel quality changes for a CR UE, it receives a new FL model adapted to the channel state.

Example results of the FL-based SS performance are presented in Figure 3. The probability of correct detection $P_d$ and probability of false alarm $P_{fa}$ (i.e., the probability of falsely detecting spectrum occupation) are considered measures of sensing quality. The PU transmission was represented by 5000 patterns per each Signal-to-Noise power Ratio (SNR) value in the form of 50-by-100 resource elements in frequency and time (thus 5000 RBs per one pattern) generated by the 5G downlink signal simulator, which mirrors the traffic-related time and frequency dependencies in the allocated RBs as in Figure 1. The received SNR has been considered in the 0–20 dB range. The shared corporate model has been built based on 8 CRs FL updates over 20 iterations by averaging the weights of their CNN models (with two hidden layers). CRs participating in FL have different channel models: 3GPP pedestrian (EPA), vehicular (EVA), and varying Doppler frequencies (in the range of 0.5 Hz – 70 Hz). The corporate model has been used and tested by three other (tester) UEs that have not taken part in the FL algorithm and have specific channel conditions and locally collected datasets.

Moreover, the tester UEs have also tested the locally learned models of individual CRs by importing them for sensing. The effects presented in Fig. 3 have been obtained by averaging the testers' results in these two sensing scenarios. The averag-ing has been done over varying Doppler frequencies of CRs (in the range of 2.5 – 55 Hz) and testers (in the range of 0.5 Hz -2.5Hz and 60 Hz – 70 Hz) over the number of CRs, from which the models are randomly imported (only for basic sensing).

Figure 3 shows that FL-based sensing performs better in building a universal model for data collected with different channel conditions than models built using data specifically for one channel type. The discrepancy in the channel conditions between CRs and testers explains this effect.

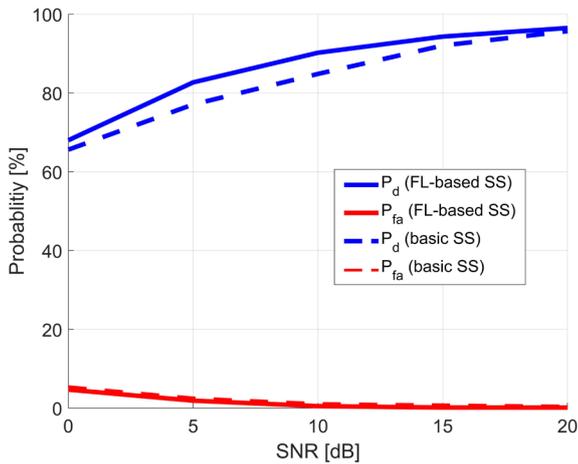

Figure 3. FL-based SS and *basic* SS performance.

To summarize, the advantages of the FL-based SS over the alternative SS schemes are the following: (i) it results in a higher quality of decisions (expressed in higher $P_d$ and similar $P_{fa}$) than autonomous (by individual sensors) sensing, including classical SS by sensors in adverse locations, and ML-based; (ii) it allows for spectrum prediction as opposed to schemes not incorporating ML such as classical SS and CS; (iii) it assures the privacy of data as opposed to centralized ML since transmission of training data is not required; and (iv) it allows for building a universal model for all SUs, ready to be used by new incoming users.

FL-based SS limitations hail from communication bottlenecks. The main challenge is decreasing the total number of communication rounds and transmitting small model updates. Moreover, the UEs participating in the training process may differ in terms of storage, computational ability, power supply, and network connectivity capabilities. Therefore, the FL SS approach must be resilient to UE failures and work with low participation. Finally, the local data set cannot be cleaned for missing values and irrelevant aspects.

### III. FL-based Sensing Security Attacks

FL-based SS involves sensing by individual CRs, ML on these devices and creating a corporate model. Therefore, the attack surface for FL-based SS includes typical attacks on sensing, attacks on individual ML algorithms, and attacks on FL.

### A. Attacks on SS in CR

Two classical spectrum sensing attacks in the context of CR are Primary User Emulation (PUE) attacks and Spectrum Sensing Data Falsification (SSDF) attacks. Both types aim to disturb the spectrum observation and users' access to the system [13]. When an attacker sends PU-like signals during the sensing time, this is referred to as a PUE attack and can prevent authorized CR access to the channels. A system's regular operation could be disrupted by malicious attackers or selfish ones that desire to use the spectrum exclusively. PUE attacks can result in bandwidth wastage, Denial of Service (DoS), connection instability, and service degradation. Identifying malicious users is crucial for protection against PUE attacks.

Sending incorrect local spectrum sensing reports to others, which causes wrong spectrum sensing decisions, is how an SSDF attack (referred to as a Byzantine attack) is launched in a cooperative (also in FL-based) SS. Attacks using SSDF are intended to reduce the probability of detection and disrupt the primary system's operations. They may also aim to increase the false-alarm probability to prevent access to spectral opportunities. Three categories of SSDF attackers can be identified: (i) a selfish SSDF (which aims to secure exclusive access to the target spectrum by deceptively reporting high PU activity), (ii) an interference SSDF (deceptively reporting low PU activity to cause a CR to interfere with the PU and other CR secondary users), and (iii) a confusing SSDF (randomly reporting true or false results on PU activity to prevent CRs from reaching consensus on the spectrum occupation).

A Generative-Adversarial Network (GAN) is an approach to generative modeling using deep learning methods that can create fake examples statistically representative of training data without having access to the client's confidential data (FL-nodes). Their operation is based on two sub-models—the generator model, which is trained to create new examples (which could potentially be considered as belonging to the original dataset), and the discriminator model, which tries to categorize these examples as either real (from the domain) or fake (generated). The two models are trained together in a zero-sum game until the discriminator model is tricked about half the time, meaning the generator model generates plausible samples [5][8]. Thus, GANs can be considered an intelligent method of PUE or SSDF.

Most contemporary defense strategies against attacks on SS are those that make direct judgments based on the most recent data on spectrum sensing and the reputation of the sensors [13].

## B. Attacks on ML

On the one hand, ML can help manage the CR network operation, e.g., by streamlining SS in the considered area; on the other, it opens the access network to a new kind of attack. The possible risks brought on by the use of ML in communication networks may be roughly categorized into two groups: ML used to develop sophisticated assaults and ML as a target for attacks aimed at lowering the security and efficiency of ML algorithms used in operating networks. The latter type of attack, which is the main focus of this paper, is designed to cause ML systems to learn wrong models, make erroneous decisions, make false predictions, or reveal confidential information. Attacks of this kind can be exploratory if they target the inference phase and are causal if they target the learning phase (training, model development). Depending on whether the attacker has complete, partial, or no knowledge of the training data, the training method, and its parameters, these attacks can be run in a white-box, gray-box, or black-box setting. The main types of attacks on ML algorithms are as follows: (i) poisoning attacks, (ii) evasion attacks, and (iii) inference attacks [8]. A study of techniques used to deceive or mislead an ML model is called Adversarial Machine Learning (AML). AML can be used to attack or crash a machine learning system. It can also be used to defend against sophisticated adversaries that utilize AI/ML algorithms to damage a system [15].

The aim of *poisoning attacks* is to influence the learning outcomes by manipulating the data or the learning algorithm in the model development phase, i.e., in the training phase. The need for new model learning based on new data makes this attack attractive to attackers since it offers them a chance to influence the trained model through data injection, data manipulation, and logic corruption (a corruption of an algorithm or its learning logic).

An *evasion attack* is aimed at the inference stage based on the previously learned model. The attacker tries to bypass the model in the test phase by introducing small perturbations in the input values. An example of an evasion attack is the generation of a signal that mimics the transmission of an authorized user at the authentication stage.

As a service in modern networks, ML algorithms are susceptible to new attacks via Application Programming Interfaces (APIs). These kinds of attacks are called *inference attacks* (also called reverse engineering) and include (i) model inversion attacks, (ii) model extraction attacks, and (iii) membership inference attacks. An inversion attack aims to recover training data or their labels using the ML algorithm results. A model extraction (stealing) attack focuses on constructing a stolen (or surrogate) model replicating the functionality or the victim model's performance. The stolen model may have a different architecture than the victim model. It is based on the observation of the results of the prediction or the time of its implementation. A membership inference attack determines whether a sample was used to train a target model by observing the model's results.

## C. Attacks on FL

Apart from the typical threats on ML, specific attacks can be observed in FL due to communication and collective operations to create an aggregated model. Moreover, attackers can take advantage of FL's design benefit: local privacy that prevents the FL server from seeing the agents' local data or training procedures. On the other hand, the collective operation of the legitimate CRs participating in FL may dominate the FL model creation process and prevent malicious nodes from negatively impacting the model. Let us then give an overview of the attacks specific to FL-based sensing.

*Poisoning attacks* targeting a subset of FL nodes can be launched as local models are retrained by freshly collected data. By embedding a well-crafted sample to data-pollute the FL process, an adversary may covertly affect the local training datasets to control the corporate model's outcome. A particular case of this attack in FL-based sensing is an SSDF attack. In this case, the training data is falsified to reflect high, low, or random PU activity and impact the spectrum occupancy model.

*Model poisoning* is related to Byzantine attacks where hostile agents can send arbitrary gradient updates to the FL server. In these situations, the adversarial objective is to induce a distributed implementation of the stochastic gradient descent algorithm to converge to completely ineffective or suboptimal models. The vulnerability of FL to adversaries that exploit the privacy these models are supposed to provide is investigated in [9].

By examining locally computed updates, *inference attacks* can extract meaningful information about the training data set or the model itself. The types of inference attacks to which FL is vulnerable include the ones described above for ML [5][10]. However, in the case of FL, they may be launched against FL nodes (CRs) or the FL server.

Usually, in FL, many communication messages must be sent back and forth between the FL server and each FL node over the iterative learning process to reach convergence. Therefore, a non-secure communication route (usually wireless) is vulnerable to *communication attacks*. For example, a Man-In-The-Middle (MITM) attack can alter the exchanged messages. A DoS or signaling storm attack would aim at the occupation of radio resources in the control channels by massive requests for access to the system. A PUE attack can also be considered a communication attack since it is based on transmitting a fake PU signal. Moreover, inference attacks are partially based on eavesdropping on the globally shared model parameters, thus requiring the decoding of encrypted messages. Additionally, communication constraints (e.g., limited bandwidth or radio resources) can undermine the FL system.

Another group of attacks involves creating fictitious local updates (e.g., using GANs) to obtain the shared global model without actually taking part in the FL process. The primary reasons for submitting false updates in *free-riding (spoofing) attacks* are to conserve local computing resources, compensate for the lack of necessary data, or avoid violating data privacy laws so that local data are unavailable for model training.

An illustration of the above-described attacks on FL-based spectrum sensing in the setup from Figure 2 is presented in Figure 4.

## IV. FL-based SS Security Measures

To improve the resistance of FL techniques to adversarial attacks, an assessment of their vulnerability is needed first, and then the application of the appropriate defense measures. Several techniques can prevent these attacks, and here below, we briefly discuss them. Moreover, several methods have been developed to counteract communication and spoofing attacks in radio access networks. It should be emphasized, however, that existing defense mechanisms resilient to attacks reviewed in the previous section are still imperfect, i.e., they cannot fully protect FL-based spectrum sensing methods for CR. The taxonomy of these attacks and countermeasures is provided in Fig. 5.

*Defenses against poisoning attacks* (including SSDF) have been put forth and surveyed in many research papers (e.g., in [11]). They can be roughly divided into *input validation* and *robust learning*. Before feeding the data into the ML model, input validation aims to clean the training (and retraining) data from malicious and anomalous samples. For instance, the *reject on negative impact* technique cleans data by eliminating examples that negatively affect learning outcomes [8]. The relevant methods first create several micro-models trained on a disjoint fraction of input samples to accomplish data cleaning. The anomalous training data subsets are then omitted by combining the micro-models in a majority voting process. In contrast to input validation, robust learning uses robust statistics to create learning algorithms resistant to contaminated training data [12].

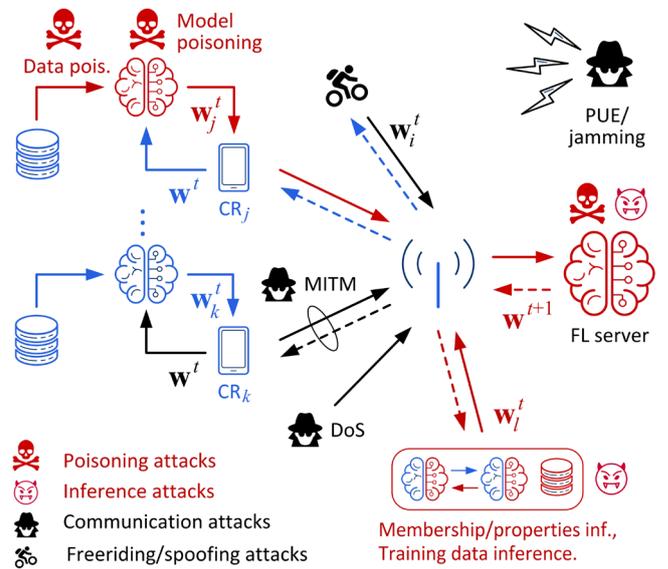

Figure 4. Illustration of FL attacks in SS.

*Defense against evasion attacks* includes adversarial training (training the model on a dataset augmented with adversarial examples), defensive distillation (training using the knowledge inferred from an ML model to strengthen its robustness), ensemble methods (combining multiple models to build a robust model), defensive GANs, and techniques to counteract the detuning of the model. By projecting input samples onto the range of the GAN's generator before feeding them into the ML model, defense GANs seek to clean them from adversarial perturbations. In other words, they seek to identify the sample that the GAN's generator can produce most similar to the adversarial example and send it as an input to the ML model.

*Defenses against inference attacks* related to the theft of the ML model and APIs of ML algorithms include the following methods: (i) learning with Differential Privacy (DP) to prevent disclosure of training data by making the model prediction independent of a single input; (ii) homomorphic encryption, which enables the model to be trained on encrypted data, thus ensuring data privacy; and (iii) limitation of sensitive information available through the API of the ML algorithm.

DP in FL-based sensing aims to ensure that no sensing record in a given FL-node dataset can be meaningfully distinguished from the other records in a highly likely scenario. This technique's primary method would be to add noise to the sensed PU's time-frequency RB occupation or the detected energy before exchanging individual updates with the FL server. The statistical data quality loss caused by the noise introduced by each FL node should be negligible in comparison to the strengthened data privacy protection. Given the required quality of SS for the efficient operation of CRs, the DP may not be practical for FL-based SS.

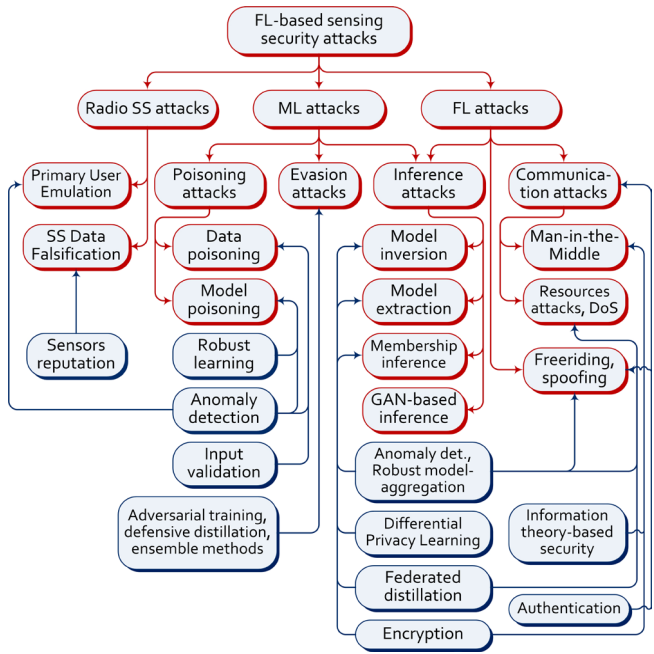

Figure 5. FL-based SS security attack and countermeasure classification.

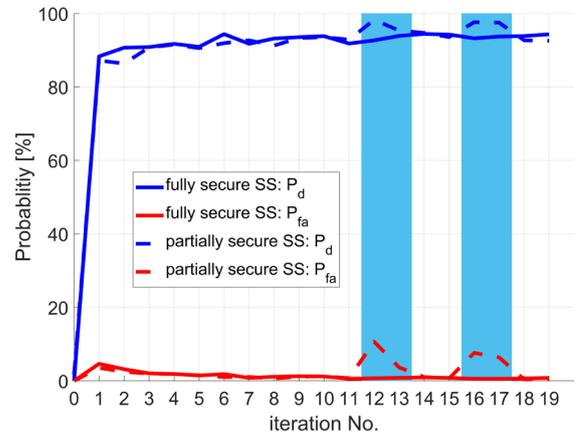

Figure 6. Secure FL-based SS performance.

The model compression technique called *federated distillation* is a method whereby a globally shared model that has received the necessary training gradually imparts the essential knowledge to a local model. The concept of disseminating information alone rather than model parameters may be used to increase security while lowering communication costs and computation overhead.

***Defenses against communication attacks***, particularly in radio-access networks, have been a research focus for many years. This is because of the open nature of the radio communication medium. A recent survey of relevant attacks and defenses in radio-access networks is found in [14]. The critical defense strategies against jamming, spoofing, eavesdropping, altering communication messages, and DoS attacks are the following: authorization and authentication procedures, data encryption, information-theory-based security algorithms, and anomaly-detection mechanisms.

***Anomaly detection methods*** can detect events that deviate from a typical pattern or activity by analytical and statistical analysis. Anomaly detection algorithms can spot problematic clients in FL environments to detect poisoning attacks, free-riders, PUE, jamming, DoS-type attacks, or incorrect model updates. If these anomalies can be identified, in some cases, they can also be eliminated.

Figure 6 presents $P_d$ and $P_{fa}$ versus the iteration number of an FL-based sensing algorithm. In our example simulation scenario, the PU signal is a 5G downlink transmission consisting of 5000 patterns in the form of 50-by-100 resource elements in frequency and time (thus 5000 RBs per pattern). The FL model is built based on 3 FL nodes: two eligible ones and an attacker poisoning the training data by incorrect labeling (50% of RBs are labeled as occupied). Moreover, one additional CR node, the tester node, tests the corporate model, although it does not participate in its creation. All nodes are characterized by randomly generated EVA channels, random Doppler frequency in the 30-55 Hz range, and a mean SNR of 10 dB. The security algorithm implemented in the FL server detects the model update anomalies using its energy-sensing dataset to test the received models. Suppose the decisions on the spectrum occupation using a particular model do not exceed a set percentage threshold of accordance with the decisions using other models. In this case, the model is rejected, i.e., it does not participate in creating a corporate model. In Figure 6, the solid lines represent the fully protected FL SS resulting from adopting a relatively high decisions-accordance threshold of 65%. The results represented by the dashed lines have been obtained for a less protective algorithm with a threshold of 55%. Note that the less-protective algorithm accepted the attacker's model in iterations 12, 13, 16, and 17, which increased $P_{fa}$. Thus, the spectrum opportunities were lost. The secure algorithm banned the attacker's model.

## V. Conclusions

FL-based spectrum sensing is characterized by higher reliability than autonomous sensing. It allows for spectrum prediction as opposed to schemes not incorporating ML techniques. It also assures the privacy of local data since only the transmission of local model parameters is required. Finally, it allows for building a corporate spectrum occupation model ready to be used by the new incoming users. However, FL-based SS can be a target of cyberattacks. The security threats originate from vulnerabilities of the applied ML and FL algorithms and the ubiquitous nature of the radio communication medium. In this paper, we have summarized potential attacks on FL-based SS and indicated methods to detect, analyze and defend against them. We have provided a taxonomy of attacks and defense methods.

Despite the capabilities of the defense methods against the attacks on FL-based SS discussed in this paper, each has its limitations, and none of them can be a one-stop-shopping solution to combat all threats. Thus, given the potential of FL for spectrum sensing in cognitive radio, robust security mechanisms are of considerable interest for future CR systems. Here, we have presented some promising results of poisoning attack detection and defense against them in FL-based SS.

Further research should be focused on balancing the power of FL-based SS and local data privacy benefits vs. wireless communication limitations and diverse capabilities of FL agents (CRs). Security metrics, such as the attack detection and mitigation probabilities, should be studied for different local learning and FL algorithms to establish security standards for their application.

**Acknowledgment**

This work was funded by the National Centre for Research and Development in Poland within the 5GStar project CYBERSECIDENT/487845/IV/NCBR/2021 on "Advanced methods and techniques for identification and counteracting cyberattacks on 5G access network and applications".

**Małgorzata Wasilewska** is a Ph.D. student and a teaching assistant at the Institute of Radiocommunications at Poznan University of Technology (PUT). Her main fields of interest are spectrum sensing, security in wireless networks, artificial intelligence algorithms, and distributed learning.

**Hanna Bogucka** is a professor at the Institute of Radiocommunications at PUT. She is involved in research in wireless cognitive and green communication. She is a member of the Polish Academy of Sciences. She also serves as the Member-at-Large of the IEEE Communications Society Board of Governors and Europe Regional Chair in the IEEE ComSoc Fog/Edge Industry Community.

**H. Vincent Poor** is the Michael Henry Strater University Professor at Princeton University, where his interests include wireless networks, energy systems, and related fields. He is a member of the U.S. NAE and NAS and a foreign member of the Royal Society and other academies. He received the IEEE Bell Medal in 2017.